\documentclass[twocolumn]{aastex62}

\usepackage{graphics,graphicx,xspace,natbib,amssymb}
\usepackage[caption=false]{subfig}
\usepackage{amsmath}
\usepackage{threeparttable}
\usepackage{makecell}

\usepackage{multirow}
\usepackage{fontawesome5}

\newcommand{\sersic}{S\'{e}rsic\xspace}

\newcommand{\comment}[1]{}

\let\itAA\AA
\renewcommand{\AA}{\mathrm{\itAA}}

\shorttitle{Minor merger growth in action}
\shortauthors{Suess et al.}

\begin{document}

\title{Minor merger growth in action:\\ JWST detects faint blue companions around massive quiescent galaxies at $0.5 \le z \le 3$}
\shortauthors{Suess et al.}

\author[0000-0002-1714-1905]{Katherine A. Suess}
\affiliation{Department of Astronomy and Astrophysics, University of California, Santa Cruz, 1156 High Street, Santa Cruz, CA 95064 USA}
\affiliation{Kavli Institute for Particle Astrophysics and Cosmology and Department of Physics, Stanford University, Stanford, CA 94305, USA}

\author[0000-0003-2919-7495]{Christina C.\ Williams}
\affiliation{NSF’s National Optical-Infrared Astronomy Research Laboratory, 950 North Cherry Avenue, Tucson, AZ 85719, USA}

\author[0000-0002-4271-0364]{Brant Robertson}
\affiliation{Department of Astronomy and Astrophysics, University of California, Santa Cruz, 1156 High Street, Santa Cruz, CA 95064, USA}

\author[0000-0001-7673-2257]{Zhiyuan Ji}
\affiliation{Steward Observatory, University of Arizona, 933 N. Cherry Avenue, Tucson, AZ 85721, USA}

\author[0000-0002-9280-7594]{Benjamin D.\ Johnson}
\affiliation{Center for Astrophysics $|$ Harvard \& Smithsonian, 60 Garden St., Cambridge MA 02138 USA}

\author[0000-0002-7524-374X]{Erica Nelson}
\affiliation{Department for Astrophysical and Planetary Science, University of Colorado, Boulder, CO 80309, USA}

\author[0000-0002-8909-8782]{Stacey Alberts}
\affiliation{Steward Observatory, University of Arizona, 933 N. Cherry Avenue, Tucson, AZ 85721, USA}

\author[0000-0003-4565-8239]{Kevin Hainline}
\affiliation{Steward Observatory, University of Arizona, 933 N. Cherry Avenue, Tucson, AZ 85721, USA}

\author[0000-0003-2388-8172]{Francesco D'Eugenio}
\affiliation{Kavli Institute for Cosmology, University of Cambridge, Madingley Road, Cambridge, CB3 0HA, UK} 
\affiliation{Cavendish Laboratory, University of Cambridge, 19 JJ Thomson Avenue, Cambridge, CB3 0HE, UK}

\author[0000-0003-4891-0794]{Hannah \"Ubler}
\affiliation{Kavli Institute for Cosmology, University of Cambridge, Madingley Road, Cambridge, CB3 0HA, UK}
\affiliation{Cavendish Laboratory, University of Cambridge, 19 JJ Thomson Avenue, Cambridge, CB3 0HE, UK}

\author[0000-0002-7893-6170]{Marcia Rieke}
\affiliation{Steward Observatory, University of Arizona, 933 N. Cherry Avenue, Tucson, AZ 85721, USA}

\author[0000-0003-2303-6519]{George Rieke}
\affiliation{Steward Observatory and Dept of Planetary Sciences, University of Arizona 933 N. Cherry Avenue Tucson AZ 85721, USA}

\author[0000-0002-8651-9879]{Andrew J.\ Bunker }
\affiliation{Department of Physics, University of Oxford, Denys Wilkinson Building, Keble Road, Oxford OX1 3RH, UK}

\author[0000-0002-6719-380X]{Stefano Carniani}
\affiliation{Scuola Normale Superiore, Piazza dei Cavalieri 7, I-56126 Pisa, Italy}

\author[0000-0003-3458-2275]{Stephane Charlot}
\affiliation{Sorbonne Universit\'e, CNRS, UMR 7095, Institut d'Astrophysique de Paris, 98 bis bd Arago, 75014 Paris, France}

\author[0000-0002-2929-3121]{Daniel J.\ Eisenstein}
\affiliation{Center for Astrophysics $|$ Harvard \& Smithsonian, 60 Garden St., Cambridge MA 02138 USA}

\author[0000-0002-4985-3819]{Roberto Maiolino}
\affiliation{Kavli Institute for Cosmology, University of Cambridge, Madingley Road, Cambridge, CB3 0HA, UK}
\affiliation{Cavendish Laboratory - Astrophysics Group, University of Cambridge, 19 JJ Thomson Avenue, Cambridge, CB3 0HE, UK} 
\affiliation{Department of Physics and Astronomy, University College London, Gower Street, London WC1E 6BT, UK}

\author[0000-0001-6106-5172]{Daniel P.\ Stark}
\affiliation{Steward Observatory, University of Arizona, 933 N. Cherry Avenue, Tucson, AZ 85721, USA}

\author[0000-0002-8224-4505]{Sandro Tacchella}
\affiliation{Kavli Institute for Cosmology, University of Cambridge, Madingley Road, Cambridge, CB3 0HA, UK}
\affiliation{Cavendish Laboratory, University of Cambridge, 19 JJ Thomson Avenue, Cambridge, CB3 0HE, UK}

\author[0000-0002-4201-7367]{Chris Willott}
\affiliation{NRC Herzberg, 5071 West Saanich Rd, Victoria, BC V9E 2E7, Canada}

\email{suess@ucsc.edu}

\begin{abstract}
Minor mergers are thought to drive the structural evolution of massive quiescent galaxies; however, existing HST imaging is primarily sensitive to stellar mass ratios $\gtrsim$1:10.  
Here, we report the discovery of a large population of low-mass companions within 35~kpc of known $\log{M_*/M_\odot}\gtrsim10.5$ quiescent galaxies at $0.5\le z\le3$. While massive companions like those identified by HST are rare, JWST imaging from JADES reveals that the average massive quiescent galaxy hosts $\sim$5 nearby companions with stellar mass ratios $<$1:10. Despite a median stellar mass ratio of just 1:900, these tiny companions are so numerous that they represent at least 30\% of the total mass being added to quiescent galaxies via minor mergers. While relatively massive companions have colors similar to their hosts, companions with mass ratios $<$1:10 typically have bluer colors and lower mass-to-light ratios than their host galaxies at similar radii. The accretion of these tiny companions is likely to drive evolution in the color gradients and stellar population properties of the host galaxies. Our results suggest that the well-established ``minor merger growth" model for quiescent galaxies extends down to very low mass ratios of  $\lesssim$1:100, and demonstrates the power of JWST to constrain both the spatially-resolved properties of massive galaxies and the properties of low-mass companions beyond the local universe.
\end{abstract}


\section{Introduction}

The {\it Hubble Space Telescope} (HST) revealed that distant quiescent galaxies appear to be remarkably compact: the sizes of passive galaxies double between $z\sim2$ and $z\sim0$ \citep[e.g.,][]{daddi05,vandokkum08,damjanov09}. The strength of color gradients in these galaxies also evolves rapidly: at $z\sim2$, quiescent galaxies are red throughout; by $z\lesssim0.5$ their outskirts are bluer than their centers \citep[e.g.,][]{suess19a,suess19b,suess20,mosleh20,miller23}.  
The major driver of this structural evolution is thought to be gas-poor minor mergers, which are effective at increasing galaxy sizes and building up color gradients while leaving central densities and stellar masses mostly unchanged \citep[e.g.,][]{bezanson09,hopkins09,naab09,ji22}. HST imaging indicates that roughly half of the observed size growth of quiescent galaxies is explainable by minor mergers with mass ratios $\ge$1:10 \citep[e.g.,][]{newman12,belli15}.

Despite their importance to quiescent galaxy evolution, the faint nature of low-mass companions means that they are difficult to characterize beyond the local universe. 
Even when limiting to the massive quiescent hosts ($\log{M_*/M_\odot}\gtrsim10.7$), HST studies such as \citet{newman12} only probed mass ratios of $\sim$1:10 at $z\sim1.5$. In a hierarchical growth scenario, smaller mergers are likely to be even more common \citep[e.g.,][]{fakhouri10}. 
The existence of lower-mass companions at $z\gtrsim0.5$ as well as their integrated effect on the masses, sizes, and color profiles of their massive quiescent hosts remains uncharacterized.

JWST \citep{gardner23} opens the door to studying faint, low-mass satellite galaxies outside of the local universe. The JWST Advanced Deep Extragalactic Survey \citep[JADES;][]{eisenstein23} reaches $\sim1.5\mu$m depths comparable to HST's Ultra Deep Field (UDF) over much larger area --- {\it and} triples the resolution, increasing the effective surface brightness of compact companions \citep{rieke23_nircam}. The redder imaging enabled by JWST/NIRCam also allows us to begin disentangling stellar population properties to understand why color gradients evolve \citep[e.g.,][]{miller22}.

In this Letter, we leverage the depth and resolution of new JWST imaging from the JADES survey to study the occurrence of low-mass companions around a sample of 161 massive quiescent galaxies at $0.5 \le z \le 3$. We find that these companions are extraordinarily common, with each galaxy in our sample hosting $\sim$5 companions with $\log{M_*/M_\odot}\le9$. 
We also investigate how these newly-discovered companions may influence the growth and evolution of their quiescent hosts by comparing the colors and stellar population properties of the companions to the spatially-resolved properties of their hosts. 

Throughout this Letter we assume a standard $\Lambda$CDM cosmology with $\Omega_m = 0.3$, $\Omega_\Lambda=0.7$, and $h=0.7$. Magnitudes are quoted in the AB system \citep{oke74}.

\section{Methods}

\begin{figure*}
    \centering
    \includegraphics[width=.95\textwidth]{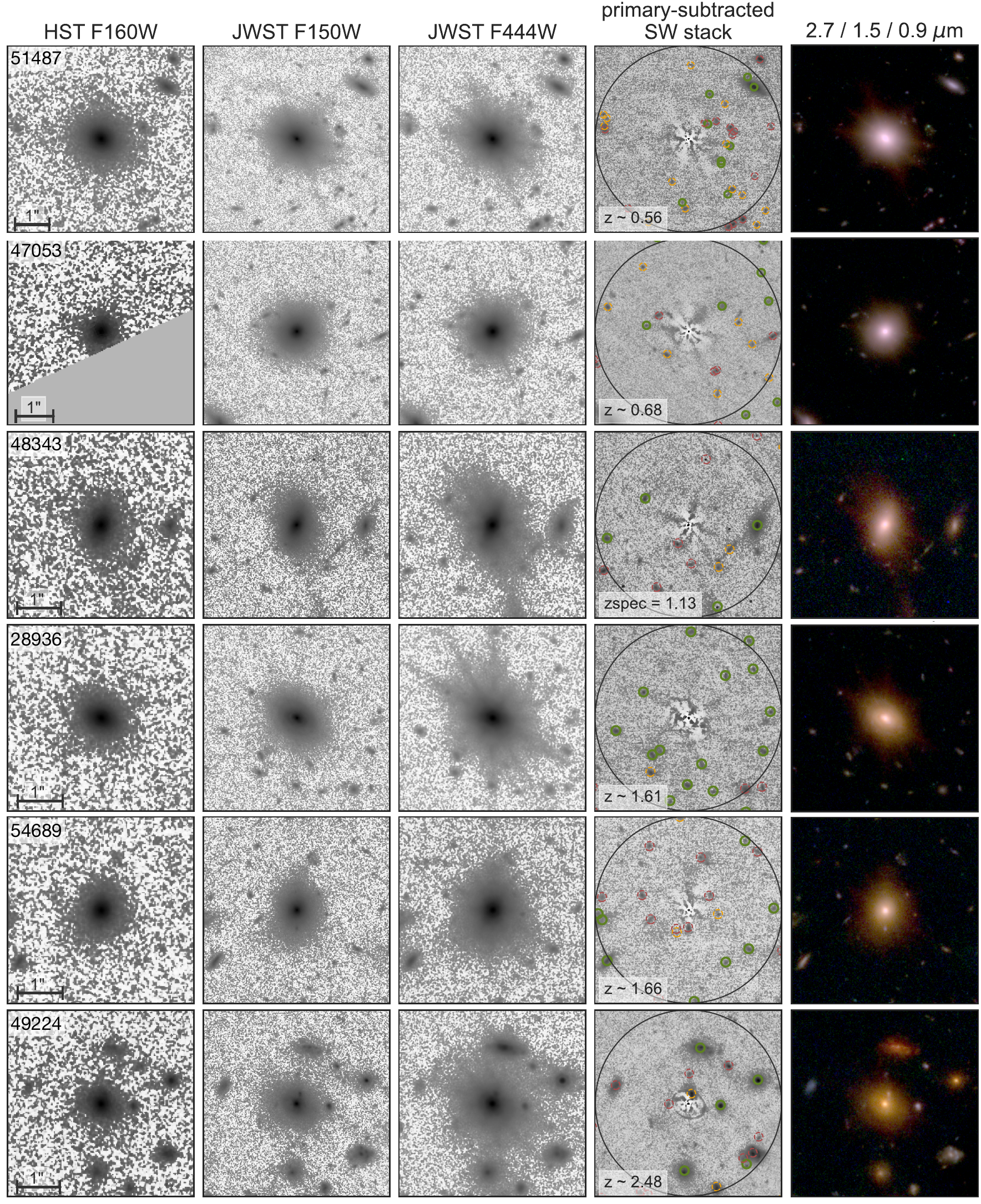}
    \caption{HST F160W, JWST F150W, JWST F444W, a primary-subtracted short-wavelength stacked JWST image, and a color JWST image. Rows are ordered by redshift. In the SW stack, potential companions with photometric redshifts consistent within 1$\sigma$ are indicated with 0.2" green circles; 2$\sigma$ with dashed 0.2" yellow circles; and inconsistent with dashed 0.2" red circles. The black circle indicates our companion search radius (35~kpc).}
    \label{fig:images}
\end{figure*}

\subsection{Data \& parent sample}
\label{sec:data}
This work uses multi-band JWST/NIRCam mosaics in GOODS-S. The data reduction is described in detail in \citet{rieke23}. We use F090W, F115W, F182M, F200W, F210M, F277W, F335M, F356W, F410M, F444W, F460M, and F480M mosaics, which include data from JADES \citep[][PID~1180]{eisenstein23}, JEMS \citep[][PID~1963]{williams23}, and FRESCO \citep[][PID~1895]{oesch23}.
We additionally use HST data in F435W, F606W, F775W, F814W, F850LP, F105W, and F160W \citep{illingworth16,whitaker19a} which were reduced and mosaicked onto the same scale as the JADES JWST mosaics. 

Our sample of 161 quiescent galaxies is described in detail in Ji et al. (in prep.). Quiescent galaxies are selected from the 3D-HST catalog \citep{skelton14} to have $\log{M_*/M_\odot}\ge10$, $z\ge0.5$, and rest-frame $UVJ$ colors within the \citet{schreiber15} quiescent box. Suspected AGN are removed by cross-matching with previous works such as \citet{lyu22}. Seven additional spectroscopically-confirmed quiescent galaxies from \citet{carnall20} are also included.  Figure~\ref{fig:buddyNumbers} shows the sample in mass-redshift space. 13 sources lie in the UDF, while the remaining 148 have HST imaging through CANDELS-deep; this means that for the majority of galaxies in our sample, JADES JWST/F150W imaging is 1-2~mag deeper as well as three times higher resolution than available HST/F160W imaging.

\subsection{Faint source detection}
We noticed many faint sources around these known quiescent galaxies that did not appear in previous HST imaging (Fig.~\ref{fig:images}). These sources are typically small, close to their host, and often brighter in bluer wavelengths. Due to a combination of color and resolution -- which impacts both blending and segmentation -- these objects are not reliably detected in the main JADES catalog, which is based on a long-wavelength stacked detection image and optimized to detect faint isolated red sources.

We therefore create a source catalog based on a short-wavelength stack of the F090W, F115W, F182M, F200W, and F210M filters. The higher resolution of this SW stack allows us to more reliably detect faint blue sources close to bright hosts. We use \texttt{sep} \citep{barbary16}, the python implementation of SourceExtractor \citep{bertin96}, to detect and deblend sources using a 3~pixel gaussian kernel, a minimum area of five pixels, a threshold of 5$\sigma$, 32 deblending levels, and a contrast of 0.001. These parameters avoid shredding our primary galaxies while still detecting and deblending most visible companions.  

\subsection{Faint source photometry}
Measuring accurate fluxes for faint sources close to bright primaries requires careful subtraction of the extended primary light profile.  
We model and subtract the light of each primary quiescent galaxy using the \texttt{photutils.isophote} elliptical isophote fitting algorithm, which requires minimal assumptions about the galaxy's intrinsic light profile \citep{jedrzejewski87}. We create a 7x7" cutout around each primary galaxy in all available JWST/NIRCam and HST/ACS bands, mask all nearby sources, then fit an ellipse to each band independently. 
As an initial guess, we use the central position, Kron radius, and axis ratio from our SW-detection catalog. We fit for both ellipse geometry and flux at $r<1.75$"; at $r>1.75$" we fix the ellipse geometry and fit only for the flux. We constrain the maximum semimajor axis to 3.5" to ensure all ellipses fit within our cutout. Some primaries are too faint in bluer bands for the ellipse fits to converge; in these cases we mask the primary using our SW-detected segmap (expanded by 11~pixels to capture galaxy outskirts). 

We use \texttt{photutils} to perform aperture photometry on the primary-subtracted cutouts at the locations of all other SW-detected sources. We use a 0.2" diameter aperture, which encloses the majority of the flux in these compact sources while minimizing background flux.    
Following \citet{rieke23}, we correct these fluxes for the PSF by applying a linear factor based on the WebbPSF encircled energy and calculate photometric errors using empty apertures. To allow for additional flux uncertainty due to our primary subtraction, we assume that each source is in the 90\% noisiest region of the mosaic independent of its actual weight map value. 

Due to our SW detection, $\sim5-10\%$ of cutouts contain a shredded star-forming galaxy at a similar photometric redshift as the primary galaxy. We remove these multiple detections of single objects by identifying any locations where $\ge$3 sources are within 0.2" of each other, then removing all sources except the one with the brightest F150W flux.

\subsection{Identification \& properties of companions}

We fit the multi-band photometry for all faint sources and primary galaxies with \texttt{eazy-py} \citep{brammer08,eazypygithub} using the {\texttt{tweak\_fsps\_QSF\_12\_v3}} template set and a free photometric redshift $0.01\le z\le12.0$. 
Faint sources are identified as possible companions if (a) they are within 35~kpc of the primary galaxy, (b) their primary-subtracted fluxes are brighter than $m_{\rm{F150W}} = 29.5$ (e.g., $\gtrsim6\sigma$ detections), and (c) their photometric redshifts from \texttt{eazy-py} are consistent with the primary quiescent galaxy within 1$\sigma$ (determined using the \texttt{eazy-py} `z160' and `z840' columns). The limit of 35~kpc ensures we can compute accurate $M/L$ gradients for the primary galaxies, while the magnitude cut ensures sources are visible above the extended host light. Our redshift cut may preferentially include sources with particularly uncertain photometric redshift solutions; our selected companions have a median $\delta z/z\sim0.4$, higher than the primary galaxy median $\delta z/z\sim0.1$. From regions free of primary sources we estimate $\lesssim15$\% of identified companions are chance alignments.  
This companion identification is unlikely to be either pure or complete; we leave a detailed selection and characterization of the full companion population around massive quiescent galaxies to future works.

We use \texttt{eazy-py} to re-fit the photometry of each possible companion with the redshift fixed to that of the primary quiescent galaxy to ensure consistent rest-frame color estimates. 
We also fit each companion with FAST \citep{kriek09} using the \citet{bruzual03} templates, the \citet{kriek13} dust law, a delayed-$\tau$ star formation history, photometric redshift fixed to that of the primary, and a grid spacing of 0.1 in A$_v$ and $\log$(age). We test fits with both free and fixed metallicity; we find no systematic differences in either stellar mass or $M/L$ with different metallicity assumptions, but allowing sub-solar metallicities for the companions --- expected based on their low stellar masses, e.g. \citet{sanders20} --- typically increases their inferred ages. 
Throughout this work, we quote stellar population properties from FAST and rest-frame colors from \texttt{eazy-py}.

\subsection{M/L profiles of primary quiescent galaxies}
We calculate radial $M/L$ profiles for the primary quiescent galaxies following the third method described in \citet{suess19a}, based on \citet{szomoru10}. In brief, we measure deconvolved surface brightness profiles for each galaxy in each band, then perform SED fitting using the same setup described above to obtain a stellar mass, age, and dust attenuation value at each radius. $M/L$ values are obtained by dividing the FAST stellar mass and the \texttt{eazy-py} rest-frame V-band light. 

We use \texttt{Lenstronomy} \citep{birrer18,birrer21} to fit a single \sersic profile to each primary quiescent galaxy in each available JWST and HST band. We use our SW-detected source catalog and segmentation map to simultaneously fit all sources that are up to two magnitudes fainter than the primary galaxy and have centers $<$2" away; we mask all other sources. Most candidate companions are significantly fainter than their primary, and are masked. 
We first fit each primary in F150W--- which offers the best combination of signal-to-noise and resolution--- allowing the position, \sersic index, half-light radius, position angle, and axis ratio to vary. We constrain the \sersic index to $0.5\le n \le 8$, the size to $0.01" \le r_e \le 10"$, and the position to within $0.15"$ of the SW catalog center. We then fit a \sersic profile to all other JWST and HST bands, forcing the center to the best-fit F150W center but allowing free $n$, size, axis ratio, and position angle. All fits are performed on the $\sim$0.03" drizzled images and use empirical point spread functions (PSFs) from \citet{ji23}. 

We measure the residual flux in concentric elliptical annuli that are centered at the primary's best-fit F150W location, have the same axis ratio and position angle as the best-fit F150W \sersic profile, and are spaced approximately one PSF FWHM apart. We add this residual flux profile to the best-fit deconvolved \sersic profile. We estimate error bars on the resulting residual-corrected deconvolved surface brightness profiles using the empty aperture scaling relations from \citet{rieke23}.  
We fit these spatially-resolved multi-band surface brightness profiles with \texttt{eazy-py} and FAST as described above, fixing the redshifts to the best-fit spatially-integrated redshift. 

\section{Results}

In Figure~\ref{fig:images}, we show example HST and JWST images of six quiescent galaxies in our sample. While JWST/F150W and HST/F160W trace nearly identical rest-frame wavelengths, JWST imaging is three times higher resolution and up to two magnitudes deeper than HST. 
Figure~\ref{fig:images} also shows a stack of all JWST short-wavelength filters after subtracting the primary quiescent galaxy. The smooth component of the primary galaxy is typically well-fit; small residuals often tracing the PSF diffraction spikes can occasionally be seen in the innermost $\sim0.5$". Possible companions are circled in green. 
These examples demonstrate the striking number of newly-revealed faint companions around these well-studied massive quiescent galaxies. Across our sample of 161 quiescent primaries, we identify a total of 629 companions, of which only 60/629 (10\%) have $\log{M_*/M_\odot}\ge9$ and 36/629 (6\%) have mass ratios $\ge$1:10.

\begin{figure}
    \centering
    \includegraphics[width=.4\textwidth]{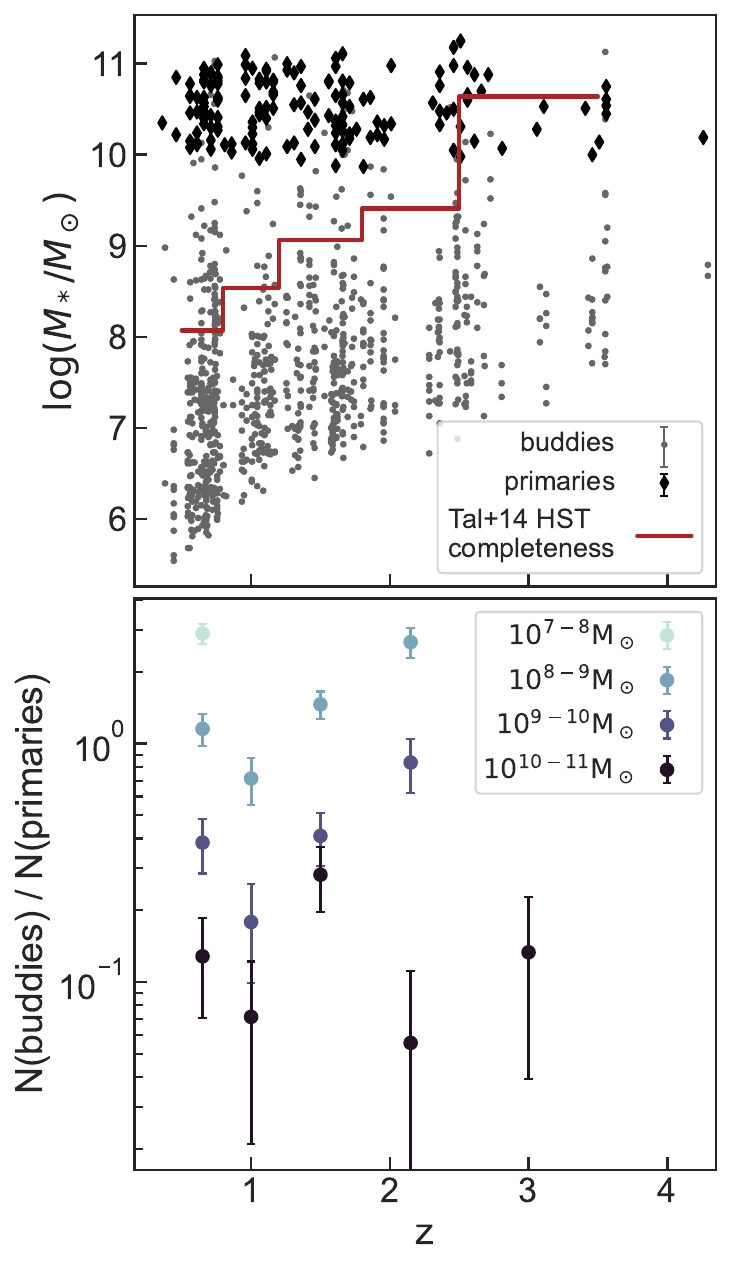}
    \caption{Top: stellar mass versus redshift for all companions (``buddies") and primary hosts; typical error bars shown in legend. The \citet{tal14} CANDELS-wide mass completeness limit is shown in red; most of our detected companions are below this limit. Bottom: number of companions per host as a function of redshift, in bins of companion mass. Points are excluded if the companion mass is less than 1.5~dex below the HST mass completeness. 
    Massive companions are rare across redshift, existing around only $\sim10$\% of hosts \citep[consistent with][]{newman12}. Low-mass companions are increasingly more common, with most massive quiescent galaxies hosting at least 1-2 companions with $8<\log{M_*/M_\odot}<9$.}
    \label{fig:buddyNumbers}
\end{figure}

Figure~\ref{fig:buddyNumbers} shows stellar mass versus redshift for our primary quiescent galaxies and all identified companions, along with the mass completeness limit of CANDELS-wide from \citet{tal14} (CANDELS-deep, which contains most of our sources, is about a magnitude deeper than CANDELS-wide in GOODS-S). The majority of companions are well below the \citet{tal14} mass completeness threshold. 
The right panel shows the number of companions per host in bins of companion mass. While $\log{M_*/M_\odot}\gtrsim9$ companions exist around just $\sim$10\% of hosts, most massive quiescent galaxies host a total of $\sim$5 $\log{M_*/M_\odot}\lesssim9$ companions. The masses of most companions are larger than expected for globular clusters; however, some of our lowest-mass sources at $z\lesssim1$ may be consistent with the high-mass end of the globular population \citep[e.g.,][]{barmby07,caldwell11}.  

\begin{figure}
    \centering
    \includegraphics[width=.48\textwidth]{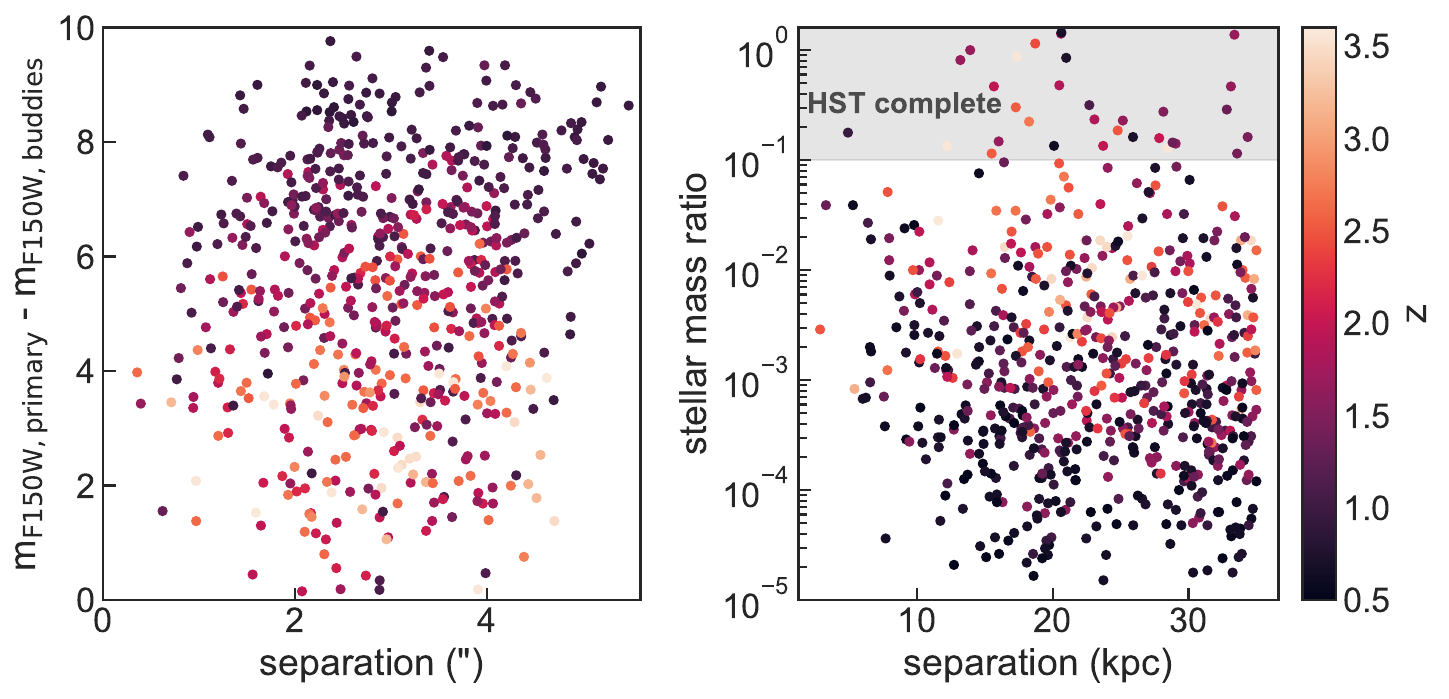}
    \caption{Left: difference in F150W magnitude between primary quiescent galaxy and all companions as a function of separation, colored by primary redshift. It is more difficult to detect faint companions that are close to their hosts, especially at high redshift. Right: inferred mass ratio between primary quiescent galaxy and companions as a function of separation. The grey shaded region indicates mass ratios $>$1:10, where previous studies such as \citet{newman12} using HST CANDELS-wide imaging were complete. Just 6\% of companions in this study were in the regime probed by HST.}
    \label{fig:massDiff}
\end{figure}

Figure~\ref{fig:massDiff} demonstrates {\it why} these companions with consistent photometric redshifts were not identified in previous studies: they are simply too faint to distinguish from the extended light of their host at HST's resolution. The grey shaded region in the plot shows the 1:10 stellar mass ratio where previous HST studies \citep[e.g.,][]{newman12} were complete; just 6\% of our companions lie in this $\ge$1:10 regime. 
Whether parameterized in terms of a difference in magnitude or inferred mass ratio, the companions are up to 10,000 times fainter than their host. While this difference seems extreme, we note again that our quiescent primaries are as bright as $m_{\rm{F150W}}\sim18.4$ and we only include companions with primary-subtracted fluxes above the 6$\sigma$ limiting depth of JADES F150W imaging: even these faint sources are easily distinguishable by eye in single-band JADES imaging (Fig.~\ref{fig:images}). 

There are clear trends with mass and redshift in Figures~\ref{fig:buddyNumbers} \& \ref{fig:massDiff}: it is easier to detect faint, nearby companions at lower redshifts. This indicates that our current companion selection is incomplete. Figure~\ref{fig:massDiff} also shows there may be a lack of bright companions at very close separations, perhaps because such massive companions would trigger starbursts and prevent the primary galaxy from entering our quiescent selection. Completeness aside, it is clear from this initial study that JWST is revealing an extraordinarily prominent population of faint, low-mass companions around bright quiescent galaxies at cosmic noon. 

We find that the sum of the stellar mass contained in all massive companions (mass ratio $\ge$1:10) is $10^{11.4}M_\odot$, while the sum of the stellar mass contained in all less-massive companions (mass ratio $<$1:10) is $10^{11.0}M_\odot$. Given that our sample is likely to be less complete at more extreme mass ratios, our results indicate that these newly-discovered low-mass companions contribute $\ge30$\% of the total mass accreted onto massive quiescent galaxies via mergers. 

\begin{figure}
    \centering 
    \includegraphics[width=.48\textwidth]{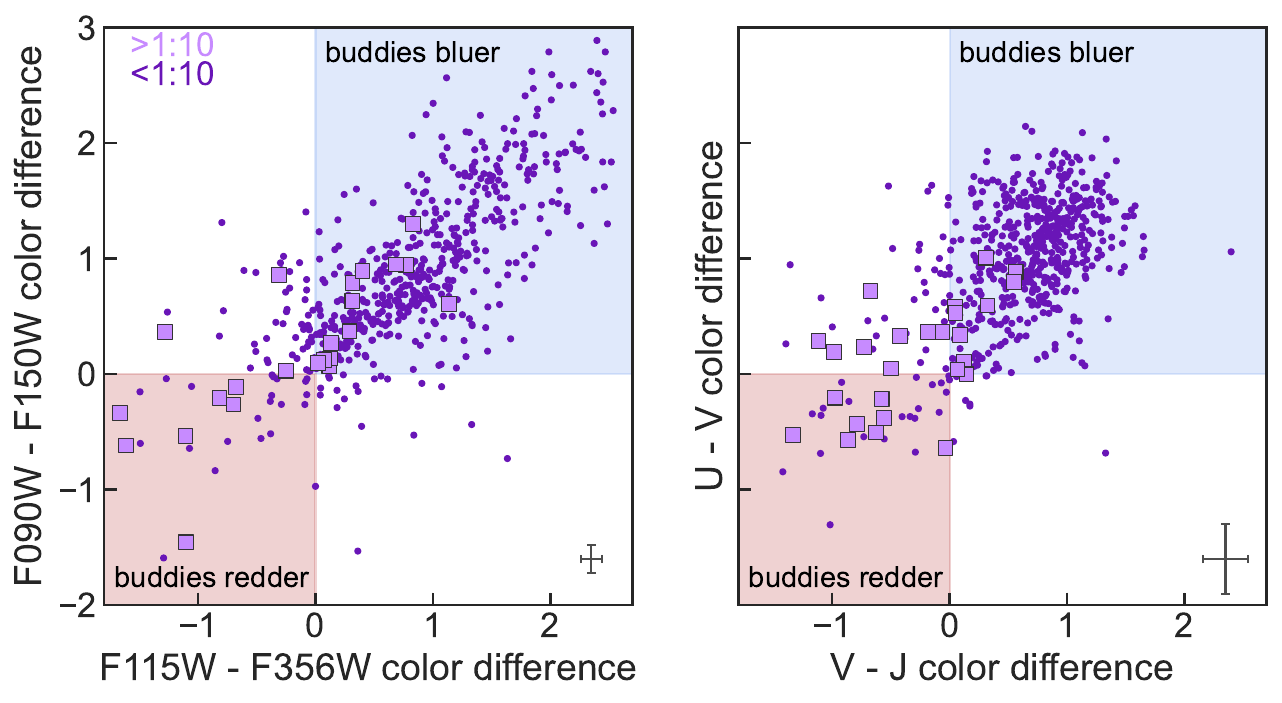}
    \caption{Difference in observed (left) and inferred rest-frame (right) color between each primary quiescent galaxy and its possible companions. 
    Most companions are bluer than their primary galaxy in both observed and inferred rest-frame color: 83\% (13\%) of companions have observed colors bluer (redder) than their primary, and 85\% (5\%) of companions have rest-frame colors bluer (redder) than their primary.}
    \label{fig:colorDiff}
\end{figure}

Figure~\ref{fig:colorDiff} shows the difference in both observed and inferred rest-frame color between each primary and all of its identified companions. Overall, most companions in our sample are bluer than their hosts, whether we consider observed or rest-frame colors: 85\% of companions have bluer $U-V$ and $V-J$ rest-frame colors than their hosts. However, there is a clear dependence on companion mass ratio: companions with mass ratios $>$1:10 are typically redder than those with lower mass ratios, with colors similar to or even redder than their hosts.

\begin{figure*}
    \centering
    \includegraphics[width=.95\textwidth]{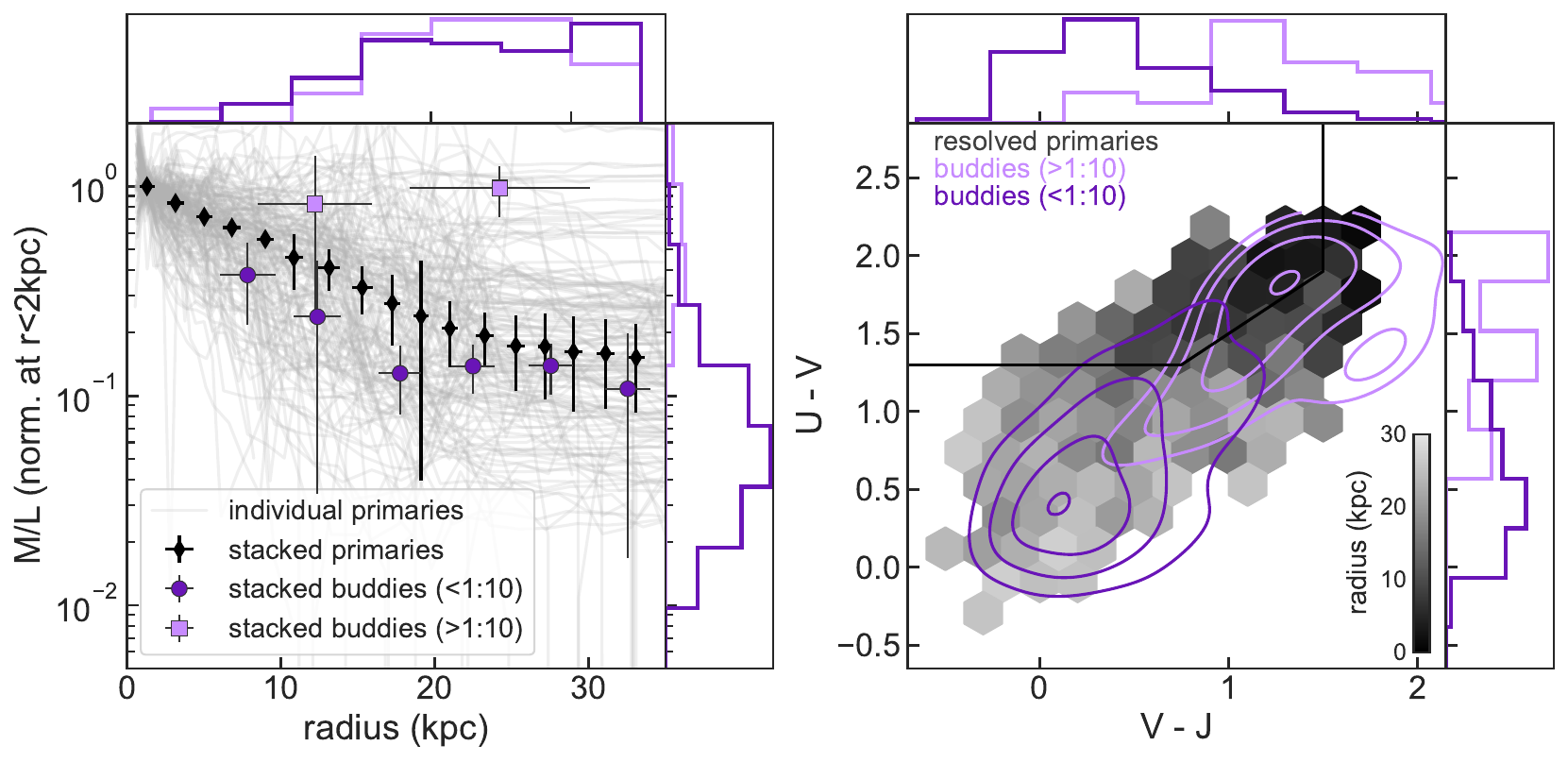}
    \caption{Spatially-resolved $M/L$ and UVJ diagrams. Left: $M/L$ profiles for each individual primary quiescent galaxy (grey curves), a median stack of all quiescent galaxies (black diamonds), and a median stack of all identified companions (purple points split by mass ratio), binned by their separation from their host. Right: Shaded hexagons show the spatially-resolved rest-frame $UVJ$ colors of all quiescent primaries; purple contours indicate the $UVJ$ colors of all companions (split by mass ratio). While individual quiescent primaries show a range of behaviors, on average most quiescent galaxies in our sample show radially decreasing $M/L$ profiles and have outskirts which lie outside the quiescent $UVJ$ box. Massive companions have relatively high $M/L$ and quiescent $UVJ$ colors, while low-mass companions have lower $M/L$ values and lie in the low-mass star-forming region of UVJ space.}
    \label{fig:ml}
\end{figure*}

In Figure~\ref{fig:ml}, we place these companions in context by comparing their properties to the spatially-resolved properties of their hosts.  The left panel of Figure~\ref{fig:ml} shows the $M/L$ profile of each quiescent galaxy as well as a stack of all 161 quiescent galaxies. While individual galaxies show a range of behaviors, the stacked profile clearly shows that radially decreasing $M/L$ profiles are common in quiescent galaxies in this mass and redshift range \citep[consistent with][]{suess19a,suess19b,suess20}. The shaded hexagons in the right panel of Figure~\ref{fig:ml} show the typical $UVJ$ colors of the primary quiescent galaxies as a function of radius. Consistent with \citet{miller22}, most galaxies in our sample are $UVJ$-quiescent in their centers but move to bluer colors consistent with star-forming galaxies in their outskirts. Our stellar population fits indicate that the lower $M/L$ values and bluer colors on the outskirts of quiescent galaxies are likely due to either younger stellar ages or lower metallicities, since most primaries have A$_v<0.5$ at all radii. 

The purple points in the left panel of Figure~\ref{fig:ml} show the companion $M/L$ values, split by mass ratio and stacked in radial bins based on their location with respect to their primary. The purple contours in the right panel show the companion $UVJ$ colors split by mass ratio. Companions with mass ratios $>$1:10 tend to have $M/L$ values similar to the center of the primary galaxy and populate the redder portion of the $UVJ$ diagram; their stellar population fits indicate a low median specific star formation rate of $\sim10^{-10.5}\rm{yr}^{-1}$. In contrast, companions with mass ratios $<$1:10 -- the new region probed by this study -- tend to have slightly lower $M/L$ than their hosts at fixed radius, and often lie in the lower-left corner of $UVJ$ space typical for low-mass star-forming galaxies. Low-mass companions located at smaller radii appear to have slightly higher $M/L$. Our stellar population fits indicate that these low-mass ratio companions also have higher median specific star formation rates of $\sim10^{-8.8}\rm{yr}^{-1}$.  
We emphasize that the companions are masked when we calculate the radial color profiles of the primary galaxies: even the smooth component of the quiescent primaries matches the low $M/L$ of the low-mass companions at large radii.

\section{Discussion \& Conclusions}

In this Letter, we have shown that most massive quiescent galaxies at $z>0.5$ have decreasing $M/L$ gradients and are surrounded by low-mass companions. The median mass ratio of our detected companions is just 1:900, well below the 1:10 limit probed by HST. These tiny companions are an order of magnitude more numerous than their higher-mass counterparts: we find that the average quiescent galaxy hosts just $\sim0.1$ companions with $10<\log{M_*/M_\odot}<11$, but $\sim1-2$ companions with $8<\log{M_*/M_\odot}<9$.   
Companions with mass ratios $>$1:10 tend to be relatively red, with $M/L$ values similar to that of their hosts; however, companions with mass ratios $<$1:10 tend to be bluer and lower $M/L$ than either the spatially-integrated or spatially-resolved properties of their primary host. Our results suggest that larger companions ($>$1:10) likely drive the majority of quiescent size growth, while tiny companions ($<$1:10) are the primary drivers of color gradient evolution. While not designed to be mass-complete, this first-look study demonstrates the power of JWST to push the study of low-mass companions to high redshifts. 

The relatively red colors, high $M/L$ values, and low ongoing SFRs of companions with mass ratios $>$1:10 are consistent with a scenario where quiescent galaxies grow via dry minor mergers \citep[e.g.,][]{bezanson09}. These red companion galaxies are unlikely to bring in significant molecular gas reservoirs that would trigger star formation in their hosts. The low median SFR of this population is also consistent with the ``conformity" picture, where quiescent galaxies tend to host quiescent satellites \citep{weinmann06}. 

New to our study is the observation that the minor merger growth model extends to very low masses: 94\% of the companions in our study have mass ratios $<$1:10. While these low-mass companions show a range in properties, the population as a whole tends to be bluer, lower $M/L$, and more highly star-forming than their host. These companions are so low-mass that they are unlikely to bring in significant gas reservoirs, but their distinct stellar populations will establish radially decreasing color gradients that likely strengthen over time as more minor mergers occur \citep[consistent with][]{suess19b,suess20}. Our observation that the outskirts of quiescent galaxies tend to be lower $M/L$ than their centers even when masking all identified companions indicates that the accretion of tiny blue companions may already have been taking place for some time before $z\sim1$. 
We find that these tiny companions make up $\gtrsim$30\% of the stellar mass accreted onto quiescent galaxies, indicating that this newly-observed population will meaningfully affect the evolution of their hosts. 
The discovery of such a large stellar mass contained in tiny companions is not inconsistent with cosmological simulations: \citet{jiang14} predict a factor of $\sim$1.5 more dark matter mass in low-mass subhalos than more massive ones, and \citet{rodriguez-gomez16} use the Illustris simulation to predict roughly equal stellar mass in mergers with mass ratios $0.1<\mu<0.25$ and $\mu<0.1$.  

While we have provided an initial guess at companion fractions in Figure~\ref{fig:buddyNumbers}, our current companion selection is not intended to be mass-complete.  
Especially at $z\gtrsim1.5$, it is unlikely we are detecting all companions with mass ratios $\lesssim$1:100. We may also be missing an intrinsically redder subset of companions -- while our 1-2$\mu$m stacked detection image does trace redward of the Balmer break across our full $0.5\le z\le3$ range, future studies may be able to account for the lower spatial resolution of the longer-wavelength JWST bands and identify redder companions from a 3-4$\mu$m detection image. Studying redder companions may provide additional insights into environmental quenching. Expanding to a larger search radius is also likely to identify many more companions; however, statistical background corrections will become more important. Finally, we note that our current work uses EAZY to calculate photometric redshifts and rest-frame colors, and FAST for stellar populations: future more detailed SED fitting may be able to further characterize stellar population trends both for the companions themselves and the spatially-resolved properties of the hosts. 

Our work demonstrates the power of JWST to constrain the spatially-resolved properties of distant galaxies from imaging alone. 
We confirm the findings of \citet{miller22} that, while the centers of quiescent galaxies often lie within the traditional UVJ box, the outskirts of quiescent galaxies are significantly bluer and often lie in the UVJ star-forming region (Fig.~\ref{fig:ml}). Our stellar population fits indicate that most quiescent galaxies in our sample have A$_v \lesssim 0.5$ throughout, suggesting that $M/L$ gradients in quiescent galaxies are likely caused by either age or metalliticy gradients. Future work is needed to break the age-metallicity degeneracy and understand the role of metallicity or abundance ratio gradients \citep[e.g.,][]{greene15,woo19,santucci20}.

Our work also ushers in a new era of studying low-mass companions outside of the local universe. While some studies have begun this work with HST \citep[e.g.,][]{nierenberg13,nierenberg16,ji18}, it is clear from Figures~\ref{fig:buddyNumbers} \& \ref{fig:massDiff} that deep observations with the efficiency and resolution of JWST are required to reliably detect $\log{M_*/M_\odot}\lesssim 9$ companions at $z>0.5$. The existence of low-mass companions at high redshift is not unexpected: after all, many dwarf galaxies in our own Local Group likely formed at $z\sim 2$ \citep[e.g.,][]{weisz14}; deep enough imaging should be able to reveal high-redshift dwarf galaxies shortly after their formation epoch. Further, cosmological simulations make strong predictions for the halo occupation distribution and the radial distribution of satellites around massive galaxies. Our results indicate that JWST data will be able to directly test these predictions against observations and constrain the relationship between stellar mass, subhalo distribution, and overall buildup of the stellar and dynamical mass of quiescent galaxies \citep[e.g.,][]{zahid17}.  
While our current study has focused on massive quiescent galaxies without clear evidence of AGN, future works may provide a detailed observational understanding of how the companion population depends on host properties such as stellar mass, star formation rate, or star formation history.

\acknowledgements 
KAS thanks Rachel Bezanson, Marijn Franx, Mariska Kriek, Phil Mansfield, Sedona Price, Richie Wang, \& Risa Wechsler for immensely helpful conversations. The authors acknowledge the FRESCO team led by PI Pascal Oesch \& the JEMS team led by PIs CCW, ST, \& Michael Maseda for developing their observing programs with a zero-exclusive-access period. 

S.C acknowledges support by European Union’s HE ERC Starting Grant No. 101040227 - WINGS. RM acknowledges support by the Science and Technology Facilities Council (STFC), by the ERC through Advanced Grant 695671 “QUENCH”, and by the UKRI Frontier Research grant RISEandFALL. RM also acknowledges funding from a research professorship from the Royal Society. AJB acknowledges funding from the "FirstGalaxies" Advanced Grant from the European Research Council (ERC) under the European Union’s Horizon 2020 research and innovation programme (Grant agreement No. 789056). FDE acknowledges support by the Science and Technology Facilities Council (STFC), ERC Advanced Grant 695671 ``QUENCH". H{\"U} gratefully acknowledges support by the Isaac Newton Trust and by the Kavli Foundation through a Newton-Kavli Junior Fellowship. DJE is supported as a Simons Investigator and by JWST/NIRCam contract to the University of Arizona, NAS5-02015. This research was supported in part by JWST/NIRCam contract to the University of Arizona NAS5-02015. BER acknowledges support from the NIRCam Science Team contract to the University of Arizona, NAS5-02015. The authors acknowledge use of the lux supercomputer at UC Santa Cruz, funded by NSF MRI grant AST 1828315. The research of CCW is supported by NOIRLab, which is managed by the Association of Universities for Research in Astronomy (AURA) under a cooperative agreement with the National Science Foundation.

\software{astropy \citep{astropy2013, astropy2018},  Lenstronomy \citep{birrer15,birrer18,birrer21}, 
          Seaborn \citep{waskom17}, WebbPSF \citep{perrin14}
          }

\bibliographystyle{aasjournal}
\bibliography{all}

\end{document}